\documentclass[twocolumn,aps,prc,superscriptaddress,showpacs]{revtex4}
\usepackage{amssymb}
\usepackage{amsmath}
\usepackage{graphicx}
\usepackage[normalem]{ulem}
\usepackage{multirow}
\usepackage{CJK}
\usepackage[usenames]{color}

\begin{document}
\begin{CJK*}{GBK}{}

\title{Iso-vector dipole resonance and shear viscosity in low energy heavy-ion collision}

\author{C. Q. Guo}
\affiliation{Shanghai Institute of Applied Physics, Chinese Academy of Sciences, Shanghai 201800, China}
\affiliation{University of Chinese Academy of Sciences, Beijing 100049, China}
\author{Y. G. Ma\footnote{Corresponding author: ygma@sinap.ac.cn}}
\affiliation{Shanghai Institute of Applied Physics, Chinese Academy
of Sciences, Shanghai 201800, China} 
\affiliation{ShanghaiTech University, Shanghai 200031, China}

\author{W. B. He\footnote{Present address: Institute of Modern Physics,  Fudan University, Shanghai 200433, China}}
\affiliation{Shanghai Institute of Applied Physics, Chinese Academy of Sciences, Shanghai 201800, China}

\author{X. G. Cao}
\affiliation{Shanghai Institute of Applied Physics, Chinese Academy of Sciences, Shanghai 201800, China}
\author{D. Q. Fang}
\affiliation{Shanghai Institute of Applied Physics, Chinese Academy of Sciences, Shanghai 201800, China}
\author{X. G. Deng}
\affiliation{Shanghai Institute of Applied Physics, Chinese Academy of Sciences, Shanghai 201800, China}
\author{C. L. Zhou\footnote{Present address: Shanghai United Imaging Healthcare  Co., Ltd., Shanghai 201807, China}}
\affiliation{Shanghai Institute of Applied Physics, Chinese Academy of Sciences, Shanghai 201800, China}

\date{\today}

\begin{abstract}

The ratio of shear viscosity over entropy density in low energy heavy-ion collision has been calculated by using the Green-Kubo method in the framework of an extended quantum molecular dynamics  model. After the system almost reaching a local equilibration for a head-on $^{40}$Ca+$^{100}$Mo collision, thermodynamic and transport properties are extracted. Meanwhile, iso-vector giant dipole resonance (IVGDR) of the collision system is also studied. By the Gaussian fits to the IVGDR photon spectra, the peak energies of IVGDR are extracted at different incident energies. The result shows that the IVGDR peak energy has a positive correlation with the ratio of shear viscosity over entropy density. This is a quantum effect and indicates a difference between nuclear matter and classical fluid.

\end{abstract}
\pacs{25.70.Ef, 21.65.Mn, 02.70.Ns, 24.10.Lx}

\maketitle

\section{Introduction}

Shear viscosity, as an important transport coefficient of fluid,  attracts more attention in recent years \cite{Dan,Shi,Fermi,Shen}. Some years ago, Kovtun, Son and Starinets found \cite{kss} that the ratio of shear viscosity $\eta$ over the entropy density $s$ has a low limit bound for all fluid, namely, the value
\begin{eqnarray}
\frac{\eta}{s} & \geq &\frac{1}{4\pi}
\end{eqnarray}
in certain supersymmetric gauge theories.
The value of $\frac{1}{4\pi}$ was claimed as the universal lower bound of shear viscosity over entropy density, i.e.  so-called KSS bound. The lower the $\eta/s$ is, the more ideal the fluid behaves. The analysis of the ultra-relativistic heavy ion collisions from the Relativistic Heavy Ion Collider (RHIC) seems to indicate that the strongly interacting quark-gluon matter behaves like a perfect liquid with the above ratio being close to the lower limit \cite{RHIC}. Many experimental efforts at RHIC and the Large Hadron Collider (LHC) as well as theoretical investigations have been carried out for the study of $\eta/s$ of this extreme hot and dense quark matter. Along this direction, temperature dependence of $\eta/s$ has been studied in high energy heavy ion collisions  \cite{PRL_new} by comparing with the LHC and RHIC data, where the partonic fluid is almost ideal. However, in heavy ion collisions at very low energy, nucleus behaves like a Fermi nucleonic fluid rather than partonic fluid, and the related study of $\eta/s$ of the nuclear matter is very limited. In this context, studies of the behavior of $\eta/s$ of nuclear matter at low temperature is of very interesting through low-energy heavy-ion collisions \cite{zcl,lsx,Muronga}.

Some previous studies have investigated the shear viscosity over entropy density for warm nuclear matter in various models, such as
for an equilibrated system of nucleons and fragments produced in multifragmentation within an extended statistical multifragmentation model \cite{Pal}, for an evolving system with the nuclear transport models \cite{lsx,zcl,zcl2,zcl3,FDQ,Deng} and thermal models \cite{XuJ} etc.
Studies also focus on the $\eta/s$ behaviour when the nuclear liquid gas phase transition \cite{LGPT,LGPT2,MaPRL,MaPRC} takes place where a local minimum of $\eta/s$ is found \cite{Pal,zcl,FDQ,Deng,XuJ}.

\begin{figure}[h]
\includegraphics[width=\linewidth]{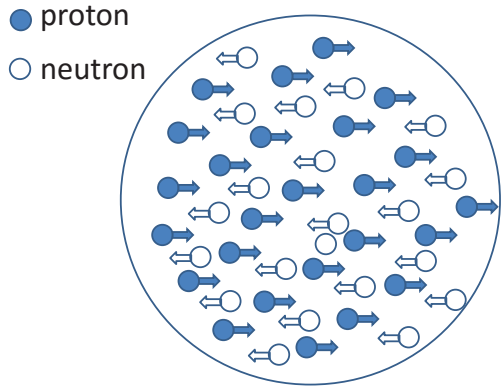}
\caption{(Color online) A schematic for the macroscopic description of iso-vector giant dipole resonance.}
\label{gdr_cartoon}
\end{figure}

Meanwhile, $\eta/s$ of low excited nuclear matter was also touched by a probe of dipole resonance in lower excitation energy region \cite{GDR,GDR2}. Fluid viscosity always plays an important role in  collective motions of fluid.
For instance, wave propagation velocity and damping are dependent on fluid viscosity.
However, wave frequency is independent on fluid viscosity.
When a wave propagates into different kinds of fluid, the frequency remains unchanged.
Giant resonance is a kind of collective motion built in nuclei and the relation between giant resonance and nuclear matter viscosity is of interesting and worth to study. On the width of giant dipole resonance, it consists of the Landau width $\Gamma^{LD}$ \cite{FIOLHAIS1986186}, the spreading width $\Gamma^\downarrow$ \cite{PhysRevLett.55.1858}, and the escape width $\Gamma^\uparrow$.
For medium and heavy nuclei, the spreading width $\Gamma^\downarrow$ gives the major contribution, which corresponds to two-nucleon interaction.
The dependence of giant dipole resonance width on shear-viscosity over entropy-density ratio has been discussed \cite{GDR2}.
In terms of macroscopic description, iso-vector giant dipole resonance (IVGDR) is considered as collective motion in which all the protons and neutrons moving together respectively with opposite phase position \cite{Goldhaber1948PR74}, as Fig.\ref{gdr_cartoon} shown.
Two components of nuclear matter, protons and neutrons, moving against each other, form a dipole oscillation which is different from classical waves in fluid.
From this classical description, the viscosity of nuclear matter should affect the frequency of IVGDR because strong viscosity may slow down  the frequency of a dipole oscillation between neutrons' centre and protons' centre.
The frequency of a dipole resonance nuclear system is represented by the peak energy of GDR.
Therefore, the viscosity may be inversely proportional to the peak energy of GDR. However, the  relation between the viscosity of nuclear matter and the  peak energy (or frequency) of iso-vector dipole oscillation is not clear so far, it deserves a detailed investigation.

Heavy ion collision is an efficient tool for investigating nuclear matter properties, from low energy to relativistic energy.
In heavy-ion fusion with low energy collision, the relation between the temperature of nuclear matter and the width of giant resonance spectra is confirmed \cite{PhysRevLett.69.249,Wang}.
In contrast,  temperature dependence of the frequency of neutron-proton oscillations indicated by the giant resonance spectra is not so clear. Of course, some experiments and model calculations show that the giant resonance spectra move to low energy when the temperature of nuclear matter gets higher \cite{Baumann1998428,PhysRevC.58.R1377}.
To discuss the viscosity dependence of the frequency of iso-vector dipole oscillation, heavy ion collisions provide an ideal venue.

In this work, an extension version of quantum molecular dynamics model  is employed to calculate $^{40}$Ca + $^{100}$Mo head-on collisions. The thermodynamic and transport properties are extracted from the nuclear fireball located in the central sphere, with the radius $R$  = 3 fm.
The rest of the paper is organizied as follows. Section \ref{sec:2} provides a brief introduction of an extended quantum molecular dynamics (EQMD) model, formula of thermodynamic properties, shear viscosity by the Green-Kubo method as well as GDR spectrum.  The relation of shear viscosity and GDR peak energies is also presented in this section. Finally a summary is given in Section  \ref{sec:3}.

\section{\label{sec:2} MODEL AND RESULTS}

\subsection{An extended quantum molecular dynamics model}

The quantum molecular dynamics (QMD) model \cite{zcl41,zcl42} approach is a many-body theory describing heavy-ion collisions in ten to GeV per nucleon range.
Later on an extension version of quantum molecular dynamics (EQMD) model,  in which the width of Gaussian wave packets for each nucleon is independent and treated as a dynamical variable   \cite{Maruyama,Cao}.
Furthermore, the Pauli potential is employed in EQMD model \cite{Maruyama} and play an important role to describe some special structures such as $\alpha$-clustering in light nuclei \cite{HeWB} which is a current hot topic in nuclear structure physics \cite{Cluster1,Cluster2,Cluster3}.
In the EQMD,
each nucleon in  a colliding system is described as a Gaussian wave packet
\begin{eqnarray}
\phi_i(\textbf{r}_i)=(\frac{\nu_i+{\nu}^*_i}{2\pi})^{3/4}exp[-\frac{\nu_i}{2}(\textbf{r}_i-\textbf{R}_i)^{2}+\frac{i}{\hbar}{\textbf{P}_i}\cdot{\textbf{r}_i}].
\end{eqnarray}
Here $\textbf{R}_i$ and $\textbf{P}_i$ are the central part of the wave packet in coordinate space and in momentum space, respectively. The complex Gaussian width $\nu_i$ is
\begin{eqnarray}
\nu_i=\frac{1}{\lambda_i}+i\delta_i.
\end{eqnarray}
Here, $\lambda_i$ and $\delta_i$ are the real part and imaginary part of wave packet, respectively.
The single nucleon density ($\rho_i(\textbf{r},t)$), matter density ($\rho(\textbf{r},t)$) in coordinate space and the kinetic energy density  ($\rho_k(\textbf{r},t)$) in momentum space can be calculated by the sum over all nucleons by the following equations, respectively

\begin{eqnarray}
\rho_i(\textbf{r},t)  &=& \frac{1}{(\pi{\lambda_i})^{3/2}}exp[-\frac{(\textbf{r}^2-\textbf{r}_i^2)}{\lambda_i}],\\
\rho(\textbf{r},t) &=& \sum\limits_{i=1}^{A_T+A_P}\rho_i(\textbf{r},t)
\\
\rho_k(\textbf{r},t) &=& \sum\limits_{i=1}^{A_T+A_P}\frac{\textbf{P}_i(t)^2}{2m}\rho_i(\textbf{r},t).
\end{eqnarray}
The total wave function of the system is a direct product of the Gaussian wave packets of nucleons
\begin{eqnarray}
\psi=\prod\limits_{i}\phi_i(\textbf{r}_i).
\end{eqnarray}
The Hamiltonian is written as
\begin{eqnarray}
 H&=&\langle \Psi|\sum_i-\frac{\hbar^{2}}{2m}{\nabla}^2_i-T_{cm}+H_{int}|\Psi \rangle\\
  &=&\sum_i[\frac{\textbf{P}^2_i}{2m}+\frac{3\hbar^2(1+\lambda^2_i\delta^2_i)}{4m\lambda_i}]-T_{cm}+H_{int},
\end{eqnarray}
where $T_{cm}$ and $H_{int}$ denote the spurious zero-point center-of-mass kinetic energy and the potential energy term, respectively. For the effective interaction $H_{int}$, we use Skyrme, Coulomb, symmetry and the Pauli potential, i.e.
\begin{equation}
H_{int}=H_{Skyrme}+H_{Coulomb}+H_{symmetry}+H_{Pauli}.
\end{equation}

Specifically, the Pauli potential is written as
\begin{equation}
H_{Pauli} = \frac{c_P}{2}\sum_i(f_i-f_0)^\mu\theta(f_i-f_0),
\end{equation}
where $f_i \equiv\sum_j\delta(S_i,S_j)\delta(T_i,T_j)|\langle\phi_i|\phi_j\rangle|^2,$ is the overlap of a nucleon $i$ with nucleons having the same spin and isospin, and $\theta$ is the unit step function. The coefficient $c_P$ is the strength of the Pauli potential.

\begin{figure}[h]
%\vspace{-1.5cm}
%\centerline{
\includegraphics[scale=1]{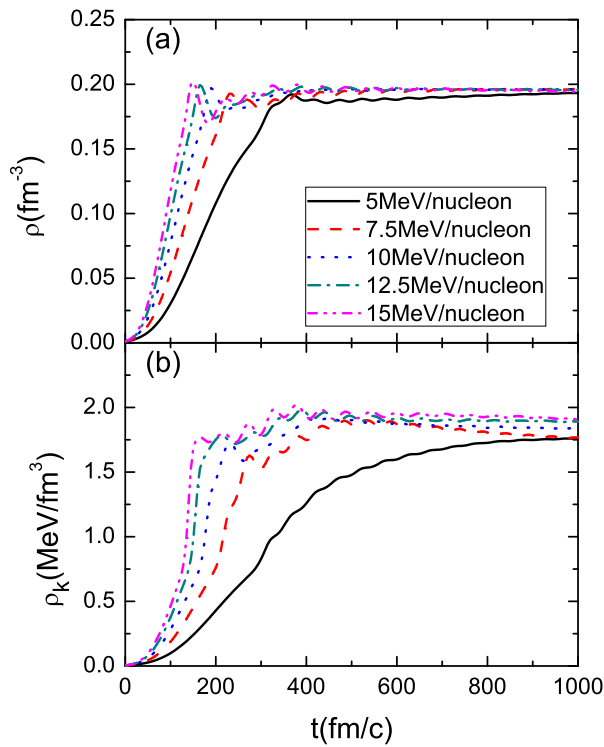}
%\vspace{-12cm}
\caption{(Color online) The time evolution of the nuclear matter density (a) and kinetic energy density (b) within a central volume ($R$ = 3 fm) for the head-on $^{40}$Ca + $^{100}$Mo collisions. Different color lines represent different incident energies.}
\label{F2}
\end{figure}

Time evolutions of the nuclear matter density and kinetic energy density in a central volume ($R$ = 3 fm) are shown in Fig.~\ref{F2}.
At different energies, both the nuclear density and kinetic energy density display rapid growth at the beginning,  then the densities of nuclear system have a small shock until they approaches equilibrium due to the  fusion(-like) reaction mechanism.

\subsection{Formula of thermodynamic properties}

Thermodynamical properties of hot nuclear matter formed in heavy ion collisions, e.g, temperature, chemical potential, and entropy density, can be extracted by different approaches.

At low temperature, $T\ll\varepsilon_F$ ($\varepsilon_F$ is the Fermi energy), the relation between the excitation energy $E^*$ and temperature $T$ is given by 
\begin{eqnarray}
E^*=aT^2.
\end{eqnarray}
For the expression of the level density the Reisdorf formalism \cite{levela} is used with a value of the parameter $a$ for $A/8$ in our calculation. When the fusion reaction is close to stable state, we assume that compound nuclear is uniform heating, so the temperature of the central region is same as the entire system.  

\begin{figure}[h]
\includegraphics[scale=1]{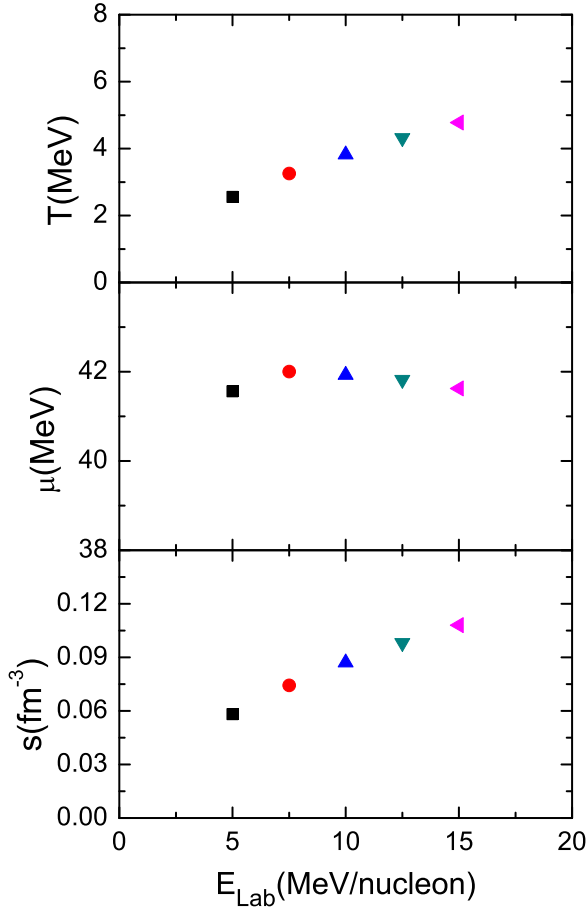}
%\includegraphics[width=1.2\textwidth]{fig3.eps}
%\vspace{-16cm}
\caption{(Color online)  The temperature (a), chemical potential (b) and entropy density (c) of the compound nuclear in a central region at different incident energies.}
\label{F3}
\end{figure}

The average temperature of the compound system in a central sphere is shown in Fig.~\ref{F3}(a). With the incident energy increasing, the nuclear temperature increases slightly.

Chemical potential $\mu_i$ of nucleon in the model can be  determined by the following implicit equation
\begin{eqnarray}
\frac{1}{2\pi^2}(\frac{2m}{\hbar^2})^{\frac{3}{2}}\int_0^{\infty}\frac{\sqrt{e_k}}{exp(\frac{e_k-\mu_i}{T})+1}de_k=\rho_i,
\end{eqnarray}
where $e_k=\frac{p^2}{2m}$ is kinetic energy and $p$ is the momentum of nucleons \cite{FDQ}.
Therefore, using this formula, one can calculate the chemical potential by nucleon's information, e.g, momentum, density, and temperature.  The chemical potential in a central sphere, when the system almost reaches a local equilibration, is shown in Fig.~\ref{F3}(b).

For the entropy density calculation,
it is straightforward to derive entropy after the density, temperature, and the chemical potential have been determined \cite{zh},
\begin{eqnarray}
S\equiv\frac{U-A}{T}=\bar{N}[\frac{5}{2}\frac{f_{5/2}(z)}{f_{3/2}(z)}-ln z],
\end{eqnarray}
where $\bar{N}$ is the number of nucleons.                $f_m(z)=\frac{1}{\Gamma(m)}\int_0^{\infty}\frac{x^{m-1}}{z^{-1}e^{x}+1}dx$ and $z=e^{\frac{\mu}{T}}$ is the fugacity.
For transforming entropy (S) and entropy density (s), we have
\begin{eqnarray}
s=\frac{\rho}{\bar{N}}S= \rho[\frac{5}{2}\frac{f_{5/2}(z)}{f_{3/2}(z)}-ln z].
\end{eqnarray}

\subsection{Ratio of shear viscosity over entropy density by Green-Kubo formula}

\begin{figure}[h]
\includegraphics[scale=1]{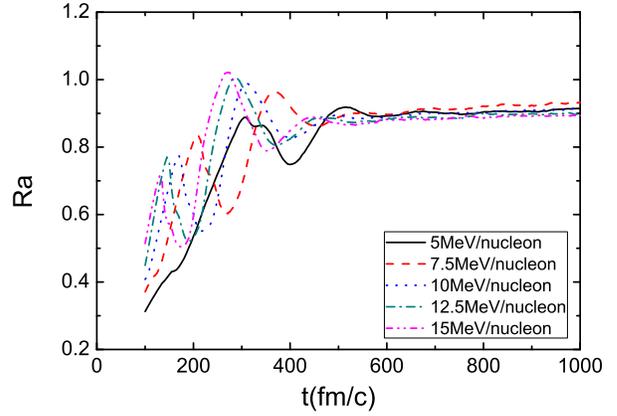}
%\vspace{-7cm}
\caption{(Color online) Time evolution of the stopping parameter in a central region at different incident energies.}
\label{F4}
\end{figure}

As mentioned above, we need to check if the equilibrium of collision system has been reached before a Green-Kubo formula can be applied. To this end, we use
a stopping parameter $R_a$ \cite{Ra}, which is defined as
\begin{eqnarray}
R_a=\sum\limits_{i=1}^{A} \frac{2\sqrt{p_x^2+p_y^2}}{\pi\sqrt{p_z^2}}
\end{eqnarray}
for checking the degree of the equilibrium.
Time evolution of $R_a$ in a central volume shown  in Fig.~\ref{F4}
illustrates that the $R_a$ approaches a saturated value close to 1, which means the nuclear system in the central region is close to equilibrium in later stage of collisions.

\begin{figure}[h]
\includegraphics[scale=0.43]{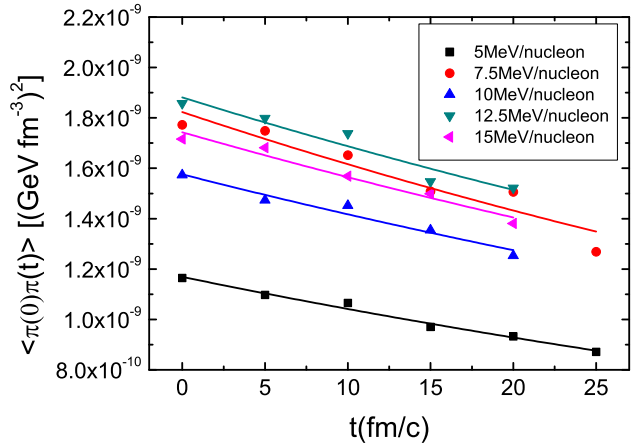}
%\vspace{-7.cm}
\caption{(Color online) The post-equilibration time evolution of the stress-tensor in a central region at different incident energies.}
\label{F5}
\end{figure}

To study the extended irreversible dynamic processes,
the Kubo fluctuation theory is employed to extract transport coefficients.
Shear viscosity determines the strength of the energy momentum fluctuation of dissipative fluxes around the equilibrium state, which can be calculated by the Green-Kubo relation.
The Green-Kubo formula \cite{gkubo} for shear viscosity is defined by
\begin{eqnarray}
\eta=\frac{1}{T}\int{d^3r}\int_0^\infty{dt} \langle \pi_{ij}(0,0)\pi_{ij}(\textbf{r},t) \rangle,
\end{eqnarray}
where $T$ is the equilibrium temperature of the system, $t$ is the post-equilibration time (`0' represents the starting time when the system tends to equilibrium), and $\langle \pi_{ij}(0,0)\pi_{ij}(\textbf{r},t)\rangle$ is the shear component of the energy momentum tensor.
The expression for the energy momentum tensor is defined by $\pi_{ij}=T_{ij}-\frac{1}{3}\delta_{ij}T_i^i$,
where the momentum tensor is written as \cite{Muronga}
\begin{eqnarray}
T_{ij}(\textbf{r},t)=\int{d^3}p\frac{p^ip^j}{p^0}f(\textbf{r},\textbf{p},t),
\end{eqnarray}
where $p^i,p^j$ is the momentum component  and $p^0$ is total energy of each nucleon, $f(\textbf{r},\textbf{p},t)$ is the phase space density of the particles.
To compute an integral, we assume that nucleons are uniformly distributed inside the volume.
Meanwhile, the spherical volume with the radius $R=3$ fm is fixed, so the viscosity becomes
\begin{eqnarray}
\eta=\frac{V}{T}\langle \pi_{ij}(0)^2\rangle \tau_{\pi},
\label{eq.23}
\end{eqnarray}
where $\tau_{\pi}$ represents relaxation time and can be extracted from the following fit
\begin{eqnarray}
\langle \pi_{ij}(0)\pi_{ij}(t)\rangle \propto{exp(-\frac{t}{\tau_{\pi}})}.
\label{eq.24}
\end{eqnarray}
As shown in Fig.~\ref{F5}, $\langle \pi_{ij}(0)\pi_{ij}(t) \rangle$ is plotted as a function of time for  $^{40}$Ca + $^{100}$Mo collision at different incident energies.
The correlation function is damped exponentially with time and can be fitted by Eq.~(\ref{eq.24}) to extract the inverse slope correspondence to the relaxation time.

\begin{figure}[h]
\includegraphics[scale=.65]{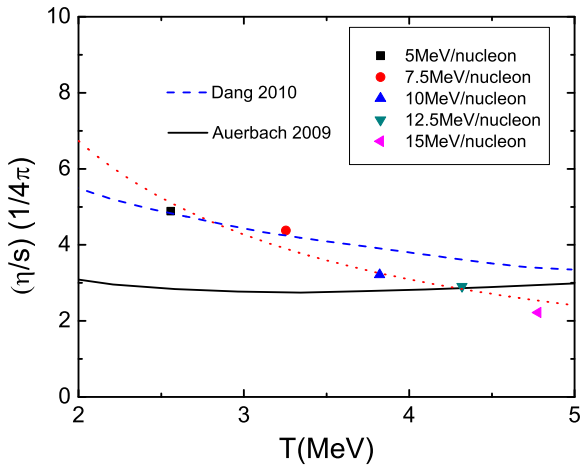}
\vspace{-7.cm}
\caption{(Color online)  The ratio of shear viscosity over entropy density as a function of temperature in a central region for the head-on $^{40}$Ca + $^{100}$Mo collisions. Solid line (Auerbach 2009) is taken from  Ref.~\cite{GDR}  and dashed line (Dang 2010) is an  extrapolation  of the phonon-damping model prediction for $^{208}$Pb in Ref.~ \cite{GDR2}.
}
\label{F6}
\end{figure}

Finally, the shear viscosity can be obtained by Eq.~(\ref{eq.23}).
Fig.~\ref{F6} shows the ratio of shear viscosity over entropy density as a function of temperature. In  low-energy $^{40}$Ca + $^{100}$Mo collisions, as the incident energy increases, the temperature slightly increases, however, the ratio of shear viscosity over entropy density of the nuclear fireball shows a slight drop. Of course, this dependence trend is consistent with our previous studies \cite{zcl,zcl2}. In the same figure, two lines are also plotted for comparison. The solid line is taken from Ref.~\cite{GDR} which is for an ideal Fermi gas, and the dashed line is an  extrapolation  of the phonon-damping model prediction for $^{208}$Pb of Ref.~ \cite{GDR2}.  Even though the systems are not the same as ours for these giant dipole resonance in  Ref.~\cite{GDR} and Ref.~ \cite{GDR2}, overall, different methods give not too much different  $\eta/s$ values.

\subsection{Giant dipole resonance}

The giant dipole resonance, that is formed during fusion in $N/Z$ asymmetry heavy-ion reactions in this paper, comes from pre-equilibrium dipole oscillations due to the charge asymmetry in the entrance channel, a so-called dynamical dipole mode. It is called pre-equilibrium GDR formed in hot nucleus, which is different from the standard GDR that is generally excited by using rapidly varying electromagnetic fields associated with photons or generated by fast electrically charged particles \cite{gdrTDHF}. For an example, the oscillation frequency of pre-equilibrium GDR is expected to be smaller because of the large deformation along the fusion path \cite{gdrTDHF,gdrbremss}.

\begin{figure}[h]
\includegraphics[scale=1]{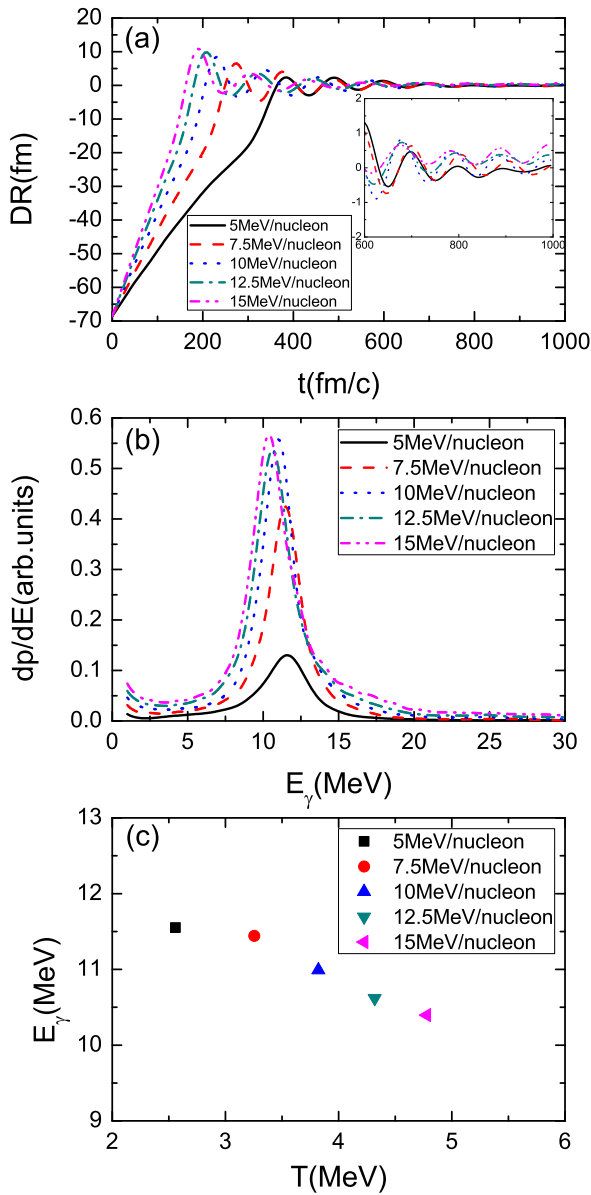}
%\includegraphics[width=1.8\linewidth]{fig7.eps}
%\vspace{-12.cm}
\caption{(Color online) Time evolution of the giant dipole moment in coordinate space (a), GDR spectra (b) and  their centroid  energies (c) for the head-on  $^{40}$Ca + $^{100}$Mo collisions with different incident energies.}
\label{F7}
\end{figure}

For a collision system, the iso-vector giant dipole moment in coordinator space DR(t)
and in momentum space DK(t) is written \cite{gdrbremss,gdr,whl,NST} as, respectively,
\begin{eqnarray}
DR(t)=\frac{NZ}{A}[R_Z(t)-R_N(t)],\\
DK(t)=\frac{NZ}{A\hbar}[\frac{P_Z(t)}{Z}-\frac{P_N(t)}{N}],
\end{eqnarray}
where $R_Z(t)$ and $R_N(t)$ are the center of mass of protons and neutrons in coordinate space, respectively; 
$P_Z(t)$ and $P_N(t)$ are the center of mass of protons and neutrons in momentum space, respectively. Fig.~\ref{F7}(a) shows the time evolution of the giant dipole oscillation in coordinate space at different incident energy. It is clear  that there are dipole oscillation at different incident energy.

Derived from the overall dipole moment D(t), one can get the $\gamma$-ray emission probability for energy $E$,  where the calculation formulas were introduced by V. Baran  et al.  in Ref.~ \cite{gdrbremss},
\begin{eqnarray}
\frac{dP}{dE} = \frac{2}{3\pi}\frac{e^2}{E\hbar{c^3}}\left|{\bar {\frac{dV_k}{dt}}(E)}\right|^2,
\end{eqnarray}
where $\frac{dP}{dE}$ can be interpreted as the average number of $\gamma$ rays emitted per energy unit, and $\frac{\bar{dV_k}}{dt}(E)$ is the Fourier transformation of the second derivative of DR(t) with respect to time,
\begin{eqnarray}
\frac{\bar{dV_k}}{dt}(E)=\int_0^{t_{max}}\frac{d^2DR_k(t)}{dt^2}e^{i(Et/\hbar)}dt.
\end{eqnarray}

By the above equation, photon emission spectrum can be obtained and shown in Fig.~\ref{F7}(b).
The results show that the frequency of dipole oscillation is dependent on temperature of nuclear matter.
Finally, the peak energies of IVGDR at different incident energies are extracted by the Gaussian fitting to the spectrum.
Fig.~\ref{F7}(c) displays that the energy of peak (centroid) positions is inversely proportional to the temperature.
This tendency is consistent with the results of previous experiments \cite{Baumann1998}.

\subsection{Relationship of GDR and $\eta/s$}

\begin{figure}[h]
\includegraphics[scale=.65]{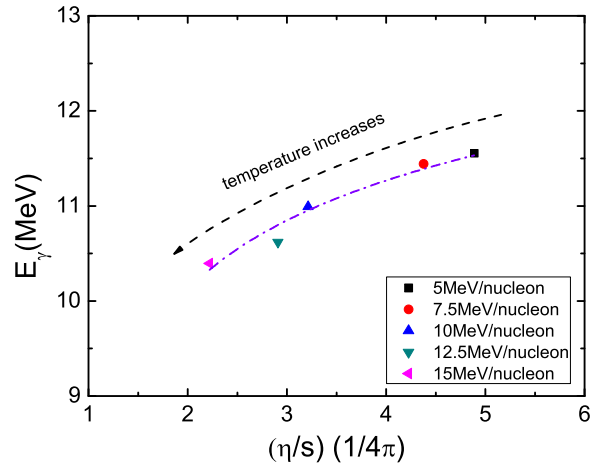}
\vspace{-7cm}
\caption{(Color online)  Peak energy of GDR as a function of the ratio of shear viscosity over entropy density for the head-on $^{40}$Ca + $^{100}$Mo collision.}
\label{F8}
\end{figure}

The relation between the ratio of shear viscosity over entropy density and peak energy is shown in Fig.~\ref{F8}.
As the temperature gets higher,  both the frequency of dipole oscillation and the shear-viscosity over entropy density becomes lower.
This is of very  interesting since the tendency is against to the prediction made from classical description as mentioned earlier in this paper.
This contradiction also indicates that the dependence between viscosity of nuclear matter and the frequency of dipole oscillation is a quantum effect, which is an important difference between nuclear matter and classical fluid.

Assuming  the above dependence between $\eta/s$ and the frequency of dipole oscillation can be extrapolated to extreme conditions of nuclear matter, one can get some interesting extrapolation. For instance,
for the extremely high temperature, where quark-gluon plasma (QGP) could be formed,  the frequency of dipole oscillation (assuming  there still exists) will be extremely low and the matter behaves like  a nearly perfect fluid with extremely low viscosity.
In other words, there is no way to get dipole excitation for a nearly perfect fluid.

\section{\label{sec:3} conclusion}

In this article, we use an EQMD model to simulate some thermodynamic quantities for a fusion system of  $^{40}$Ca + $^{100}$Mo at beam energy from 5 to 15 MeV/nucleon.  The ratio of shear viscosity over entropy density is obtained by applying the Green-Kubo formula after the fusion system is almost equilibrated, and its value is about $\frac{(2-5)}{4\pi}$ in this energy range. Meanwhile, IVGDR spectra are obtained for the system and the  peak energies are extracted at each incident energy (or temperature of the fusion system). From the Gaussian fits to the IVGDR spectra, peak energy shows a slight decrease with the increasing of temperature. By temperature dependencies of both $\eta/s$ and  peak energies of IVGDR, the relation of the peak energies of IVGDR versus the ratio of shear viscosity over entropy density of the system  is established,  and a positive correlation was found. This behaviour seems against to the guess from classical description of fluid. In this context, the $\eta/s$  dependence  of the frequency (peak energy) of dipole oscillation is a kind of quantum effect.

\begin{acknowledgments}
This work was supported in part by the
National Natural Science Foundation of China under contract Nos. 11421505, 11220101005 and 11305239, the Major State Basic Research Development Program in China under Contract No. 2014CB845401, and the  Key Research Program of Frontier Sciences of the CAS under Grant NO. QYZDJ-SSW-SLH002.

\end{acknowledgments}

\end{CJK*}

\begin{thebibliography}{99}
%1
\bibitem{Dan} P. Danielewicz, Phys. Lett. {\bf 146B}, 168 (1984).
%2
\bibitem{Shi} L. Shi and P. Danielewicz, C {\bf 68}, 064604 (2003).
%3
\bibitem{Fermi}C. Cao, E. Elliott, J. Joseph {\it et al.}, Science {\bf 331}, 58 (2011).
%4
\bibitem{Shen}C. Shen, U. Heinz, Nucl. Phys. News {\bf 25}, No. 2, 6 (2015).
%5
\bibitem{kss} P. K. Kovtun, D. T. Son, and A. O. Starinets, Phys. Rev. Lett. {\bf 94}, 111601 (2005).
%6
\bibitem{RHIC} K. Adcox, {\it et al.}, Nucl. Phys. A {\bf 757},184 (2005); B. B. Back {\it et al.}, ibid. {\bf 757}, 28 (2005); J. Arsene {\it et al.}, ibid. {\bf 757}, 1 (2005);  J. Adams {\it et al.}, ibid {\bf 757}, 102 (2005).
%7

\bibitem{PRL_new}G. Denicol, A. Monnai, and B. Schenke,  Phys. Rev. Lett. {\bf 116}, 212301 (2016).

\bibitem{zcl} C. L. Zhou, Y. G. Ma, D. Q. Fang,  G. Q. Zhang, Phys. Rev. C {\bf 88}, 024604 (2013).
%8
\bibitem{lsx} S. X. Li, D. Q. Fang, Y. G. Ma,  C. L. Zhou, Phys. Rev. C {\bf 84}, 024607 (2011); ibid,  Nucl. Sci.  Tech. {\bf 22}, 235 (2011).
%9
\bibitem{Muronga} A. Muronga, Phys. Rev. C {\bf 69}, 044901 (2004).
%10
\bibitem{Pal}S. Pal,  Phys. Rev.  C {\bf 81}, 051601(R) (2010).
%11
\bibitem{zcl2} C. L. Zhou, Y. G. Ma, D. Q. Fang, S. X. Li and G. Q. Zhang, Europhys. Lett. {\bf 98}, 66003 (2012).
%12
\bibitem{zcl3}C. L. Zhou,Y. G. Ma, D. Q. Fang, G. Q. Zhang, J. Xu, X. G. Cao, and W. Q. Shen, Phys. Rev. C {\bf 90}, 057601 (2014).
%13
\bibitem{FDQ} D. Q. Fang, Y. G. Ma, C. L. Zhou, Phys. Rev. C {\bf 89}, 047601 (2014).

\bibitem{Deng}X. G. Deng,  Y. G. Ma, and M. Vesels\'ky, Phys. Rev. C {\bf 94}, 044622 (2016).
%14
\bibitem{XuJ}  J. Xu, L. W. Chen, C. M. Ko, B. A. Li, Y. G. Ma, Phys. Lett. B {\bf 727}, 244 (2013); J. Xu, Nucl. Sci.  Tech. {\bf 24},  050514 (2013).
%15
\bibitem{LGPT}C. A. Ogilvie {\it et al.}, Phys. Rev. Lett. {\bf 67}, 1214 (1991); M. B. Tsang {\it et al.}, Phys. Rev. Lett. {\bf 71}, 1502 (1993); Y. G. Ma and W. Q. Shen, Phys. Rev. C {\bf 51}, 710 (1995).
%16
\bibitem{LGPT2}J. B. Natowitz, K. Hagel, Y. Ma, M. Murray, L. Qin, R. Wada, and J. Wang, Phys. Rev. Lett. {\bf 89}, 212701 (2002); X. Campi, J. Phys. A {\bf 19}, L917 (1986); X. Campi, Phys. Lett. B {\bf 208}, 351
(1988); A. Bonasera, M. Bruno, C. O. Dorso, and P. F. Mastinu, Riv. Nuovo Cimento {\bf 23}, 1 (2000).
%17
\bibitem{MaPRL}Y. G. Ma, Phys. Rev. Lett. {\bf 83}, 3617 (1999).
%18
\bibitem{MaPRC}Y. G. Ma, J. B. Natowitz, R. Wada {\it et al.}, Phys. Rev. C {\bf 71}, 054606 (2005).
%19
\bibitem{GDR}N. Auerbach, S. Shlomo, Phys. Rev. Lett. {\bf 103}, 172501 (2009).
%20
\bibitem{GDR2} N. D. Dang, Phys. Rev. C {\bf 84}, 034309 (2011).
%21
\bibitem{FIOLHAIS1986186} C. Fiolhais, Annals of Physics {\bf 171}, 186 (1986).
%22
\bibitem{PhysRevLett.55.1858} J. C. Bacelar, G. B. Hagemann, B. Herskind {\it et al.}, Phys. Rev. Lett. {\bf 55}, 1858 (1985).
%23
\bibitem{Goldhaber1948PR74} M. Goldhaber and E. Teller, Phys. Rev. {\bf 74}, 1046 (1948).
%24
\bibitem{PhysRevLett.69.249} G. Enders, F. D. Berg, K. Hagel {\it et al.}, Phys. Rev. Lett. {\bf 69}, 249 (1992).

\bibitem{Wang} K. Wang, Y. G. Ma, G. Q. Zhang {\it et al.}, Phys. Rev. C {\bf 95}, 014608 (2017).
%25
\bibitem{PhysRevC.58.R1377} G. Gervais, M. Thoennessen, and W. E. Ormand, Phys. Rev. C {\bf 58}, R1377(R) (1998).
%26
\bibitem{Baumann1998428} T. Baumanna, E. Ramakrishnana, A. Azharia {\it et al.}, Nucl. Phys. A {\bf 635} 428 (1998).
%27
\bibitem{zcl41} J. Aichelin, Phys. Rep. {\bf 202}, 233 (1991).
%28
\bibitem{zcl42} C. Hartnack, R. K.Puri, J. Konopka, S. A. Bass, H. St\"{o}cker, W. Greiner, Eur. Phys. J. A {\bf 1}, 151 (1998).
%29
\bibitem{Maruyama} T. Maruyama, K. Nitta, A. Iwamoto, Phys. Rev. C {\bf 53}, 297 (1996).

\bibitem{Cao}X. G. Cao, Y. G. Ma, G. Q. Zhang, H. W. Wang, A. Anastasi, F. Curciarello, V. De Leo, Journal of Physics: Conference Series, {\bf  515},  012023 (2014).

\bibitem{HeWB}W. B. He, Y. G. Ma, X. G. Cao, X. Z. Cai and G. Q. Zhang, Phys. Rev. Lett. {\bf 113}, 032506 (2014); ibid, Phys. Rev. C {\bf 94}, 014301 (2016); W. B. He, X. G. Cao, Y. G. Ma {\it et al.}, Nucl. Techniques (in Chinese) 37, 100511 (2014).

\bibitem{Cluster1} W.  von Oertzen, M.  Freer and Y. Kanada-Enyo,  Phys. Rep. {\bf 432}, 43 (2006).

\bibitem{Cluster2} H. Horiuchi, K. Ikeda and K. Kato,  Prog. Theor. Phys. Suppl.  {\bf 192}, 1( 2012).

\bibitem{Cluster3} Y. Kanada-Enyo, M. Kimura, F. Kobayashi {\it et al.}, Nucl. Sci. Tech. {\bf 26}, S20501 (2015).

%30
\bibitem{levela} W. Reisdorf {\it et al.}, Z. Phys. A {\bf 300}, 227 (1981).
%31
\bibitem{zh} H. Zheng and A. Bonasera, Phys. Rev. C {\bf 86}, 027602 (2012).
%32
\bibitem{Ra} G. Q. Zhang, Y. G. Ma, X. G. Cao, C. L. Zhou, X. Z. Cai, D. Q. Fang, W. D. Tian, H. W. Wang, Phys. Rev. C {\bf 84}, 034612 (2011).
%33
\bibitem{gkubo} R. Kubo, Rep. Prog. Phys. {\bf 29}, 255 (1966).
%34
\bibitem{gdrTDHF} C. Simenel, Ph. Chomaz, and G. de France, Phys. Rev. Lett. {\bf 86}, 2971 (2001).  

\bibitem{gdrbremss} V. Baran, D. M. Brink, M. Colonna, and M. Di Toro, Phys. Rev. Lett. {\bf 87}, 182501 (2001).

\bibitem{gdr} V. Baran, M. Cabibbo, M. Colonna {\it et al.}, Nucl. Phys. A {\bf 679}, 373 (2001).
%35
\bibitem{whl} H. L. Wu, W. D. Tian, Y. G. Ma {\it et al.}, Phys. Rev. C {\bf 81}, 047602 (2010).

\bibitem{NST}S. Q. Ye, X. Z. Cai, D. Q. Fang {\it et al.}, Nucl. Sci. Tech. {\bf 25}, 030501 (2014); C. Tao, Y. G. Ma, G. Q. Zhang {\it et al.},  Nucl. Sci. Tech. {\bf 24}, 030502 (2013).

\bibitem{Baumann1998} T. Baumann, E. Ramakrishnan, A. Azhari {\it et al.}, Nucl. Phys. A {\bf 635}, 428 (1998).

\end{thebibliography}
\end{document}